# Nuclear Magnetic Resonance Imaging with 90 nm Resolution

H.J. Mamin,[1*] M. Poggio,[1,2] C. L. Degen[1] and D. Rugar[1]

[1]*IBM Research Division*
*Almaden Research Center*
*650 Harry Rd.*
*San Jose CA 95120 USA*

[2] *Center For Probing the Nanoscale*
*Stanford University*
*476 Lomita Mall*
*Stanford, CA 94305 USA*

Magnetic resonance imaging, based on the manipulation and detection of nuclear spins, is a powerful imaging technique that typically operates on the scale of millimeters to microns. Using magnetic resonance force microscopy, we have demonstrated that magnetic resonance imaging of nuclear spins can be extended to a spatial resolution better than 100 nm. The two-dimensional imaging of $^{19}$F nuclei was done on a patterned $CaF_2$ test object, and was enabled by a detection sensitivity of roughly 1200 nuclear spins. To achieve this sensitivity, we developed high-moment magnetic tips that produced field gradients up to $1.4 \times 10^6$ T/m, and implemented a measurement protocol based on force-gradient detection of naturally occurring spin fluctuations. The resulting detection volume of less than 650 zl represents 60,000× smaller volume than previous NMR microscopy and demonstrates the feasibility of pushing magnetic resonance imaging into the nanoscale regime.

*H. Jonathon Mamin (corresponding author)
mamin@almaden.ibm.com
408-927-2502

Martino Poggio
poggiom@us.ibm.com
408-927-1547

Christian Degen
cldegen@us.ibm.com
408-927-2889

Daniel Rugar
rugar@almaden.ibm.com
408-927-2027

Magnetic resonance imaging (MRI) has had revolutionary impact on the field of non-invasive medical imaging and is finding increasing applications in material and biological sciences. Its spatial resolution, however, is of order a few micrometers at best[1-3]. Magnetic resonance force microscopy (MRFM) has been proposed as a method to overcome the sensitivity limitations of inductively detected MRI[4, 5] and push the resolution into the nanometer and, ultimately, the atomic scale[6, 7]. MRFM methods have steadily improved since the first demonstrations[8-10], with significant recent advances in both electron spin and nuclear spin detection[11-13]. Here, we demonstrate that MRFM is now capable of two-dimensional nuclear magnetic resonance imaging with 90 nm spatial resolution. In terms of resolvable volume, this work represents an improvement of 60,000× over the highest resolution conventional MRI microscopy[1] and at least 70,000× over previous MRFM-based nuclear spin imaging[14].

MRFM uses a magnetic tip and an ultrasensitive cantilever to sense the magnetic force generated between the tip and spins in a sample. Magnetic resonance is used to periodically invert the spins through application of an applied rf field at frequency $\omega_{rf}$. The rf field will affect only those spins in the vicinity of the "resonant slice," defined as the localized region where the resonance condition $\left|\mathbf{B}_{tip}(\mathbf{r}) + \mathbf{B}_{ext}\right| = \omega_{rf}/\gamma \equiv B_{res}$ is met. Here $\gamma$ is the gyromagnetic ratio, $\mathbf{B}_{ext}$ is the externally applied field, and $\mathbf{B}_{tip}$ is the field produced by the tip. For a tip that produces a highly inhomogeneous field, the effective thickness of the resonant slice will be extremely narrow. Large magnetic field gradients are therefore the key to achieving high spatial resolution. In addition, high field gradients are essential to obtaining a high signal-to-noise ratio, since the force between the magnetic tip and the spins in the sample is directly proportional to the gradient. A key enabler for the present work is the development of magnetic tips that generate magnetic field gradients as high as 1.4 million tesla per meter. Using these high gradient tips, we have demonstrated the imaging capabilities of the technique on a patterned $CaF_2$ test object.

Unlike the permanent magnet tips previously used for MRFM detection of electron spin resonance[11, 15], the present tips are based on a thin film of magnetic material that has high magnetic moment, but is magnetically soft. In particular, we use tips fabricated with sputter-deposited $Co_{70}Fe_{30}$ having magnetization $\mu_0 M = 2.3$ T. This magnetization is somewhat higher than that of iron and more than twice as high as the permanent magnetic material $SmCo_5$, which is commonly used for MRFM[11, 15]. A soft magnetic material is suitable for nuclear spin MRFM because nuclear spin experiments are typically performed in a strong magnetic field, which conveniently magnetizes the tip.

The tips we developed are compatible with "sample-on-cantilever" experiments of the type shown in Fig. 1a. In such an experiment, the sample is placed on the distal end of an ultrasensitive cantilever situated above the magnetic tip. By choosing the sample-on-cantilever configuration, rather than tip-on-cantilever, we eliminated the magnetic damping that occurs when a soft magnetic tip is vibrated in the presence of an external magnetic field[16-18].

**Results and Discussion**

<u>**Thin film magnetic tips**</u>

The magnetic tips shown in Fig. 1c and 1d were fabricated by sputtering 8 nm Fe/100 nm CoFe/10 nm Ru onto an array of micromachined silicon cones[19]. The Fe layer was present to prevent intermixing of the CoFe active magnetic layer with the silicon substrate. The Ru layer provided a protective layer against oxidation. Despite the relatively thick magnetic film, the tip remained sharp, with a radius of curvature below 20 nm. We used an array of tips to ensure that one tip was always accessible within the limited range of our piezoelectric scanner.

To quantitatively characterize the magnetic tips, we performed nanometer-scale magnetometry by measuring the MRFM signal as a function of tip-sample distance and applied magnetic field[15, 20]. The sample was a 100 nm-thick film of $CaF_2$ (99.99%) evaporated onto the end of a mass-loaded silicon cantilever (Fig. 1)[21]. Rather than rely on the thermal equilibrium (Boltzmann) spin polarization, which requires waiting on the order of a spin-lattice relaxation time $T_1$ between measurements, we used the naturally-occurring statistical polarization[22-25]. This time-dependent polarization (or spin noise) is due to fluctuations in the magnetization of the paramagnetic nuclear spins that result in a net longitudinal polarization whose root-mean-square value scales as $\sqrt{N}$, where $N$ is the number of spins in the ensemble. An rf frequency sweep method (Fig. 2 inset) was used to drive adiabatic spin inversions, thereby modulating the z-component of the nuclear magnetization at the cantilever frequency[12, 26]. The periodic reversal of the sample magnetization in the presence of the tip field gradient produces a periodic force that excites the cantilever whenever there is a naturally occurring left-right statistical imbalance in the spin polarization. Because the signal originates from the statistical spin polarization (which can be either positive or negative and has mean value that is essentially zero), the signals referred to below are based on the variance of the time-dependent cantilever amplitude.

The MRFM signal was measured as a function of external magnetic field $\mathbf{B}_{ext}$, as shown in Fig. 2a for four different values of tip-sample separation. A measurable MRFM signal is obtained whenever some portion of the resonant slice lies within the volume of the sample. The minimum value of the external field $B_{ext,onset}$ for which a signal is obtained corresponds to the field where the resonant slice barely intersects the sample surface. Since the resonant slice is defined by $B_{total} = B_{res}$, where $B_{total} = \left|\mathbf{B}_{tip}(\mathbf{r}) + \mathbf{B}_{ext}\right|$ and $B_{res}$ = 2.89 T, this implies that the tip field directly above the tip at the sample surface is given simply by $B_{res} - B_{ext,onset}$. Tip field values as a function of separation can thus be inferred from the data in Fig. 2a by determining the value of $B_{ext,onset}$ for each tip-sample separation, and can be read off the graph directly. However, because the onset of the signal as a function of $B_{ext}$ is gradual, we developed a simple computer model that was used to facilitate a more precise determination of the tip field at the sample surface (see Methods section).

The resulting tip field values are plotted as a function of the tip-sample spacing in Fig. 2b. For the closest spacings, we found that a 23 nm shift in spacing gave a 33 mT shift in tip field, resulting in a gradient $\partial B_z / \partial z$ = $1.4 \times 10^6$ T/m (14 G/nm). This is the highest magnetic field gradient from a tip that has been documented for MRFM, roughly 6× higher than that produced with rare earth tips used for electron spin detection[11, 15].

**<u>Two dimensional imaging using cyclic-CERMIT protocol</u>**

The high magnetic field gradient is a crucial part of performing nuclear MRFM imaging with high resolution. Despite the large gradient, achieving adequate signal-to-noise ratio (SNR) is still a challenge. To further enhance SNR and minimize imaging time, we explored several spin manipulation and detection protocols, in addition to the rf frequency sweep method mentioned above. The method that gave the best performance was a modification of a protocol introduced by Garner, Marohn and co-workers called CERMIT (Cantilever Enabled Readout of Magnetization Inversion Transients)[27]. A timing diagram of our version of the protocol, called cyclic-CERMIT, is shown in Fig. 3.

Cyclic-CERMIT relies on the fact that the cantilever frequency is slightly modified by the polarization of the nuclear spins near the magnetic tip. In particular, for a magnetic moment $m_z$ above the magnetic tip, the cantilever frequency is shifted by[27] $\delta f_c = \frac{1}{2} m_z \frac{f_c}{k} \frac{\partial^2 B_{total,z}}{\partial x^2}$. Frequency shifts are typically on the order of a few millihertz. To make this frequency shift distinguishable from other static frequency shifts, the spin polarization is periodically reversed by the application of a low duty cycle rf pulse which, in combination with the cantilever vibration, causes the spin polarization to invert. This periodic inversion of the spins results in a periodic modulation of the cantilever frequency, which is detected by a software-based frequency demodulator. Operationally, this protocol is nearly identical to the iOSCAR protocol developed previously, except that the rf pulse sequence is inverted[11,23,24].

Because the rf field is pulsed with a low duty cycle (~0.5%), the rf-induced heating is minimized, allowing us to operate at a temperature of 600 mK. This results in lower thermomechanical noise of the cantilever and thus helps improve the SNR.

We have combined the cyclic-CERMIT technique with the high gradient CoFe tips to perform two-dimensional nuclear magnetic resonance imaging on the 100 nm scale. To demonstrate the imaging capability, a patterned test sample was created using a focused ion beam to mill features into the end of the cantilever, which was then coated with 80 nm of $CaF_2$ as indicated in Fig. 4a. The electron micrograph in Fig. 4b shows four silicon pillars formed at the end of the cantilever with some evidence of the $CaF_2$ coating. The $CaF_2$ appears to overhang the ends of the pillars somewhat, though the contrast in the electron micrograph is low. We have made our best estimate for the dimensions of the $CaF_2$ islands based on the micrograph in Fig. 4b and other similar ones. The structure is shown schematically in Fig. 5a.

To form an image, the magnetic tip was scanned laterally in a plane at a spacing of 45 nm from the $CaF_2$ islands. This spacing was determined by contacting the sample and retracting the tip a known amount with a piezoelectric tube scanner. The cyclic-CERMIT protocol was implemented while oscillating the cantilever with an estimated peak amplitude $x_{pk}$ of 115 nm and a rf pulse width $\tau_p$ =50 μs. During the pulse the resonant slice swings a total distance of ~105 nm (peak-to-peak). With the external field adjusted to 2.83 T, spatially localized magnetic resonance signals were observed, leading to the image shown in Fig. 5c. The image is composed of an array of 60×20 pixels spaced 30 nm apart, where each pixel represents the spin-noise power after 10 minutes of averaging. To confirm that the image data were indeed due to $^{19}$F magnetic resonance, we verified that the spin signal disappeared, as expected, when the external field was raised or lowered by as little as 80 mT.

The image clearly reflects the overall morphology of the sample: the signal was strongest over the $CaF_2$ islands, and the larger gaps between the islands were well resolved. In particular, the 100 nm wide gap was resolved with nearly 100% contrast, indicating a lateral resolution better than 100 nm. This resolution is consistent with that seen in the line scan in Fig. 5d, which shows raw data taken from a cut through the image (dotted line). The slopes of the features are well fit by convolving the sample object with a simple two-dimensional (cylindrically symmetric) gaussian point spread function with a full width half maximum of 90 ± 10 nm. This resolution is comparable to the peak-to-peak motion of the resonant slice during the rf pulse $\tau_p$.

The cyclic-CERMIT protocol was modeled (see Methods section) using our estimated values for the cantilever oscillation amplitude, rf pulse width, sample dimensions, and magnetic tip parameters, all of which can affect the achieved resolution. Overall, the resulting simulated image in Fig. 5b is in reasonably good agreement with the experimental image. In particular, the 100 nm gap on the left is fully resolved in the model, as in the data. The images clearly differ in some details, however, such as the island on the right, which is barely visible in the experimental image. The discrepancy could have been caused by the resonant slice pulling out of the sample due to a tip trajectory that did not follow the tilt of the sample.

Unlike previous MRFM imaging experiments, the MRFM image did not show ring-like features associated with the resonant slice point spread function[14, 28, 29]. This is due to the extreme sharpness of the magnetic tip, which leads to a compact resonant slice with a radius of curvature that is comparable to the 100 nm sample features. We note that these magnetic tips would be better matched to imaging even smaller objects. With image deconvolution, they are capable of higher resolution, provided that the SNR is sufficient to detect the smaller volume of spins.

We can estimate the volume of material contributing to the spin signal by considering the lateral resolution and the thickness of the $CaF_2$ sample: 90 nm × 90 nm × 80 nm, or 650 zl. This volume contains roughly 30 million nuclear spins and represents a factor of 60,000× smaller volume than achieved with conventional MRI microscopy[1]. It also represents a factor of 70,000× improvement over previous MRFM-based nuclear spin imaging[14].

The above estimates assume that the entire film thickness contributes to the signal; a more detailed model calculation suggests that, under our imaging conditions, the resonant slice penetrates roughly half the film thickness and encompasses only about 10 million spins. Since the signal originates from the $\sqrt{N}$ statistical polarization, this implies that there are roughly 3200 net (rms) spins contributing to the signal. Based on simple scaling of the signal by our observed (power) SNR of 7.5, we estimate that our detection sensitivity for unity SNR is roughly 1200 nuclear spins-rms after 10 minutes of averaging.

Alternatively, we can use the equation $\delta f_c = \frac{1}{2} m_z \frac{f_c}{k} \frac{\partial^2 B_{total,z}}{\partial x^2}$ to convert the measured frequency noise to a magnetic moment noise. From a magnetostatic model for our tip geometry, we estimate the peak field derivative $\partial^2 B_{total,z}/\partial x^2$ to be of order $2\times10^{13}$ T/m$^2$ at the apex of the resonant slice, assuming a completely magnetized tip. For our 10 minute averaging time with detection bandwidth of 0.44 Hz, the rms frequency noise was found to be 1.3 mHz, yielding a detection noise floor of 200 spins-rms. The discrepancy between this estimate and the previous one is partly explained by the fact that not all the spins in the resonant slice experience the same peak field derivative. It may also suggest that some spins are not contributing at full strength, perhaps due to violation of the adiabatic condition during the spin manipulations[26].

In summary, we have shown that by combining magnetic field gradients of over $10^6$ T/m with force gradient detection of statistical spin polarization, we can extend magnetic resonance imaging into the nanoscale regime. Further improvements can be expected as MRFM techniques are further refined. For example, increasing the gradient another factor of ten, to at least $10^7$ T/m, should be possible with improved tip geometry and smaller tip-sample separation. This would lead to order-of-magnitude increases in both spatial resolution and force generated per spin. Work is also underway to develop more efficient rf field sources so as to lower overall system temperature to the low millikelvin range and thus dramatically reduce cantilever thermal vibration noise. The combination of these improvements should allow MRFM to push deeper into the nanometer regime and approach the capability needed for direct three-dimensional imaging of individual macromolecules.

**Methods**

**Experiment**

Two silicon cantilevers of the same nominal dimensions were used in the experiments. They both had resonant frequencies of $f_c \sim 3$ kHz, and spring constants $k \sim 6\times10^{-5}$ N/m, as estimated from the thermomechanical noise. The cantilever quality factor varied from 50,000 in zero applied field to 8,000 in a field of 3T. The reason for this excess magnetic dissipation is currently the subject of further investigation. The cantilever displacement was monitored with a fiber-optic interferometer.

The $CaF_2$ sample was deposited onto the end of the cantilever through thermal evaporation. The film structure consisted of 9 nm Cr/50 nm Au/100 nm $CaF_2$ for the gradient determination experiment, and 9 nm Cr/50 nm Au/80 nm $CaF_2$ for the imaging experiment. The purpose of the Cr/Au underlayer was to provide electrical screening of laser-induced charge noise in the cantilever. This underlayer was found to greatly reduce the frequency fluctuations that are observed when the cantilever is brought close to the magnetic tip. For the imaging sample, the cantilever was first shaped in a focused ion beam by making three cuts edge-on before the thin film deposition (see Fig. 4). The sample-on-cantilever configuration has practical advantages over the magnet-on-cantilever configuration, such as lower magnetic damping. It is also a natural configuration for future MRFM experiments, where ultimately the sample may be a molecular-sized object on a nanomechanical cantilever.

A 300 μm diameter copper coil was used to generate a radio frequency (rf) magnetic field at $\omega_{rf}/2\pi = 115.7$ MHz with an estimated rf field strength $B_1$ of 2 mT. For $^{19}$F spins, which have a gyromagnetic ratio $\gamma/2\pi = 40.05$ MHz/T, this rf frequency corresponds to a resonance field $B_{res} = \omega_{rf}/\gamma = 2.89$ T. The microscope was operated in vacuum at cryogenic temperatures in order to reduce the thermomechanical cantilever noise. The temperature was typically 11K for the tip calibration measurements in which the rf power was applied continuously, and 0.6 K for the cyclic CERMIT measurements.

For both the rf sweep and cyclic-CERMIT protocols, a lock-in detection scheme was used to detect the signal synchronously with the rf modulation. The detected signal is the cantilever amplitude for the rf sweep method, and the cantilever frequency shift (Fig. 3) for cyclic–CERMIT. The in-phase response contains both signal and measurement noise, while the quadrature channel represents just the measurement noise. By taking the difference in the variances, we obtain a zero-baseline signal that represents only the contribution from the spins[11].

The optimum SNR is obtained when the lock-in detection time constant is properly matched to the correlation time of the signal (i.e., the spin relaxation time)[30]. We used a bank of filters implemented in software and chose the one that gave the best SNR. For the rf sweep (field gradient) measurements, the equivalent noise bandwidth of the measurement was 1.8 Hz. For the cyclic-CERMIT measurements, the equivalent noise bandwidth was 0.44 Hz. The narrower bandwidth with cyclic-CERMIT was possible because the spins exhibited longer correlation times in this case, presumably because of the reduced duty cycle of the rf field, which was roughly 0.5%.

**Modeling**

For the rf sweep (field gradient determination) experiments, the mean square force was modeled using the following integral:

$$\left\langle F_{spin}^2 \right\rangle = A^2 \mu_N^2 \int_{vol} \eta(\mathbf{r}) \left| \frac{\partial B_{total,z}}{\partial x} \right|^2 N(\mathbf{r}) dV \quad , \tag{1}$$

where the **x** and **z** directions are defined in Fig. 1a, and the integral is taken over all space. Here $N(\mathbf{r})$ is the number of spins per unit volume in the sample and takes into account the sample geometry. $B_{total,z}$ is the **z** component of the total field, $\mu_N$ is the magnetic moment of $^{19}$F, and $A$ is a scaling factor that should in theory equal unity. The function $\eta(\mathbf{r})$ characterizes the effectiveness of the adiabatic reversals and contains the physics of the resonance condition. We approximated $\eta(\mathbf{r})$ as a binary function that equals unity when $\left| B_{total}(\mathbf{r}) - B_{res} \right| < \omega_{mod}/\gamma$, and zero otherwise; that is, $\eta(\mathbf{r})$ has unity value when the resonant slice passes through the location **r** as the rf frequency is swept from $\omega_{rf} + \omega_{mod}$ to $\omega_{rf} - \omega_{mod}$. The model assumed a uniformly magnetized spherical tip for simplicity, so that

$$B_{total,z}(x,y,z) = \sqrt{\left[\frac{3\mu_0 m_{tip} xz}{4\pi\left(x^2+y^2+z^2\right)^{5/2}}\right]^2 + \left[\frac{3\mu_0 m_{tip} yz}{4\pi\left(x^2+y^2+z^2\right)^{5/2}}\right]^2 + \left[\frac{\mu_0 m_{tip}(2z^2-x^2-y^2)}{4\pi\left(x^2+y^2+z^2\right)^{5/2}} + B_{ext}\right]^2}. \quad (2)$$

Here $m_{tip} = \frac{4}{3}\pi M_{tip} r_0^3$ is the total magnetic moment of the tip, where $r_0$ is the tip radius, and $\mu_0 M_{tip}$ was assumed equal to 2.3 T. Good fits to the experimental data were enforced by varying the tip radius and overall scaling parameter $A$ for each curve. The resulting best-fit tip radii ranged from 30 to 50 nm. Values for $A \sim 0.2$ were typical, implying that the size of the force signal was smaller than one would expect from this simplified model, which assumes idealized adiabatic reversals. Once the best-fit tip radius was determined, the field from the tip at the sample surface was easily calculated from

$$B_{tip}(0,0,r_0+d) = \frac{\mu_0}{2\pi}\frac{m_{tip}}{(r_0+d)^3},$$ where $d$ is the distance between the tip and sample surface. The resulting values were plotted in Fig. 2b.

The cyclic-CERMIT results in Fig. 5 were modeled by calculating the mean square cantilever frequency shift $\langle(\delta f_c)^2\rangle$ resulting from the spins as they were inverted by the vibrating cantilever. The simulated signal in this case was obtained from

$$\langle(\delta f_c)^2\rangle = C^2 \frac{f_c^2 \mu_N^2}{4k^2} \int_{vol} \eta(\mathbf{r}) \left|\frac{\partial^2 B_{total,z}}{\partial x^2}\right|^2 N(\mathbf{r}) dV \quad , \quad (3)$$

where $N(\mathbf{r})$ is the spin number density, $C$ is a scaling factor of order unity, and $\eta(\mathbf{r})$ is a binary function that is zero unless the moving resonant slice passes over the location **r** one time (and only one time) in moving between the extrema of its motion. The motion of the resonant slice is given by the physical motion of the cantilever during the time that the rf field is applied. In our case the rf field was applied for 30% of the time it took for the cantilever to swing from one extremum to the other, in which time the resonant slice moved the distance $2x_{pk}\sin(\pi f_c \tau_p) = 230$ nm $\cdot \sin(0.15\pi) \approx 105$ nm. For these calculations, we modeled the tip as a 100 nm-thick uniformly magnetized conical shell with a cone angle of 26° as measured by electron microscopy. This tip geometry was used to numerically compute the function $\mathbf{B}_{total}(\mathbf{r})$. For the spin density $N(\mathbf{r})$, we assumed that the sample had the idealized structure shown in Fig. 5a. With these inputs, $\langle (\delta f_c)^2 \rangle$ was then calculated as a function of tip scan position using (3). The result was the image shown in Fig. 5b.

The same tip model was used to arrive at the estimate $\dfrac{\partial^2 B_{total,z}}{\partial x^2} \sim 2 \times 10^{13}$ T/m$^2$ presented in the imaging section.

**Acknowledgements**

We thank J. Marohn for discussions on the CERMIT technique, B. Hughes for assistance with magnetic tip preparation, B.W. Chui for cantilever fabrication, and D. Pearson and B. Melior for technical support. We acknowledge support from the DARPA QUIST program administered through the US Army Research Office, the Swiss National Science Foundation, and the Stanford-IBM Center for Probing the Nanoscale, a NSF Nanoscale Science and Engineering Center.

**Author contributions**

H.J.M., D. R. and M.P. conceived, designed and performed the experiment. M.P. and D.R. implemented the rf sweep method. D.R., M.P. and H.J.M. performed tip-field modeling. C.L.D. modeled the cyclic-CERMIT protocol and performed the image simulation. H.J.M. and D.R. co-wrote the paper. All authors discussed the results and commented on the manuscript.

**Competing interests statement**

The authors declare that they have no competing financial interests.

**Correspondence** and requests for materials should be addressed to H.J.M. (mamin@almaden.ibm.com).

Fig. 1 Basic setup and components of the MRFM experiment. **a** A cantilever with a thin-film sample at the end was oriented perpendicular to a substrate supporting a conical magnetic tip. Nuclei that are within the "resonant slice" region, indicated by the dotted lines near the tip, can undergo magnetic resonance. The tip can be scanned under the cantilever using a piezoelectric actuator. **b** Single crystal silicon cantilever of the type used in the experiment. The thick end of the cantilever was coated with an evaporated thin film of $CaF_2$ to form the sample. **c** Scanning electron micrograph of an array of magnetic tips formed by coating etched silicon cones with a thin film of CoFe. The CoFe alloy generates a magnetic field with a strong gradient. **d** Close-up of an individual magnetic tip.

Fig. 2 Characterization of the tip magnetic field. **a** MRFM signal as a function of externally applied magnetic field, for different tip-sample spacings. The width of each spectrum depends upon the strength of the magnetic field produced by the tip at the stated tip-sample spacing. A slight systematic offset of $-5aN^2$ has been removed from the data. All data were acquired with a detection bandwidth of 1.8 Hz. The inset shows the rf sweeps performed twice per cantilever cycle that drive adiabatic spin inversions. **b** The inferred tip field values as a function of distance from the tip. The peak field gradient is roughly $1.4 \times 10^6$ T/m.

Fig. 3. Timing diagram for the cyclic-CERMIT technique. The cantilever is always oscillated at its natural resonant frequency. This frequency is slightly modified by the interaction between the spins in the sample and the magnetic tip, specifically the force gradient between the spins and the tip. Periodically, an rf pulse of width $\tau_p$ is applied synchronously with the cantilever motion. While the rf field is applied, the cantilever motion sweeps the resonant slice through the spins, driving adiabatic spin inversions. Each time the spins are inverted, the resulting force gradient changes sign, leading to a square wave modulation of the cantilever frequency. The shorter the pulse width $\tau_p$, the smaller the sample region where the spins are inverted. For the imaging experiments, $\tau_p$ was roughly 50 μs, or 30% of the time it took the cantilever to swing from one extremum to the other. The resulting square wave modulation in the frequency signal, which is at half the rf pulsing frequency, was detected using a software frequency demodulator.

Fig. 4. Sample preparation method for patterned $CaF_2$ test sample. **a** Close-up of the end of the cantilever shown in Fig. 1b. The end of the cantilever was first etched into a narrow finger of width 180 nm using a focused ion beam (FIB). The cantilever was then oriented edge-on into the ion beam, and three cuts were made from the side as indicated by the white lines. After the cuts were made, the $CaF_2$ film was thermally evaporated end-on. **b** Scanning electron micrograph showing the end of the cantilever after the $CaF_2$ coating. Four silicon pillars are visible with $CaF_2$ appearing to slightly overhang the gaps between the pillars.

Fig. 5 Experimental results and simulation showing two dimensional imaging of $^{19}$F nuclei. **a** Schematic of the $CaF_2$ structure used for the imaging test object. The structure represents a thin film that was evaporated onto a template etched into the end of the cantilever with a focused ion beam. All dimensions are in nanometers and are taken from electron micrographs such as the one shown in Fig. 4b. **b** Simulated image for the cyclic-CERMIT protocol using a conical tip model. **c** Magnetic resonance image taken at a tip-sample spacing of 45 nm, with contrast reflecting the spatially varying signal power from $^{19}$F nuclei. The resonant field was 2.89T, with an applied field of 2.83T. Generally good correlation with the expected morphology is observed, although the island at the right of the image is barely visible. This discrepancy is perhaps due to a slight tilt of the sample with respect to the plane of the scan. The data were acquired with a measurement bandwidth of 0.44 Hz. **d** Line scan showing raw image data taken from the location of the dotted line in **c**. The 100 nm and 250 nm gaps in the test sample are both resolved with essentially 100% contrast.

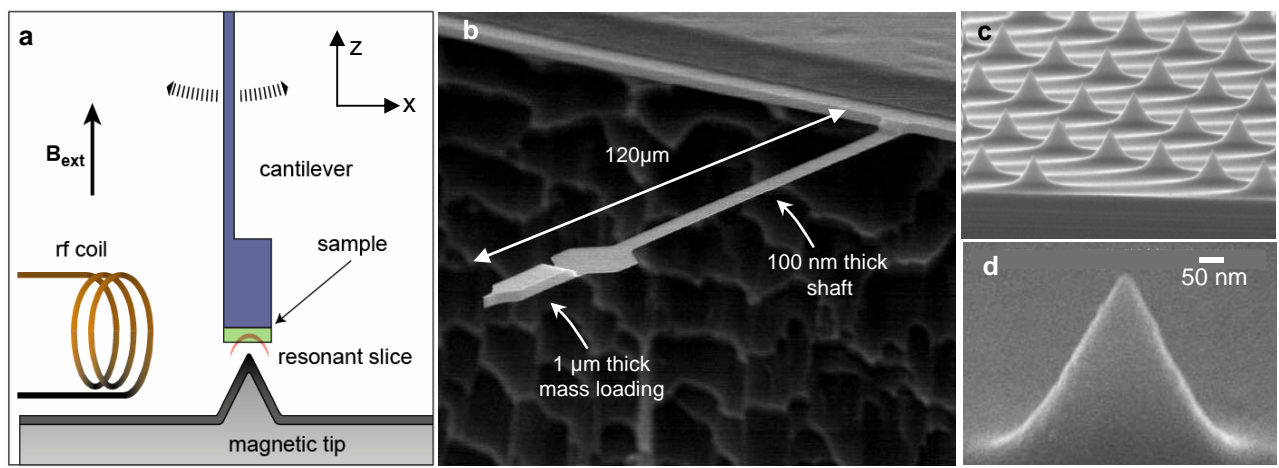

Fig. 1

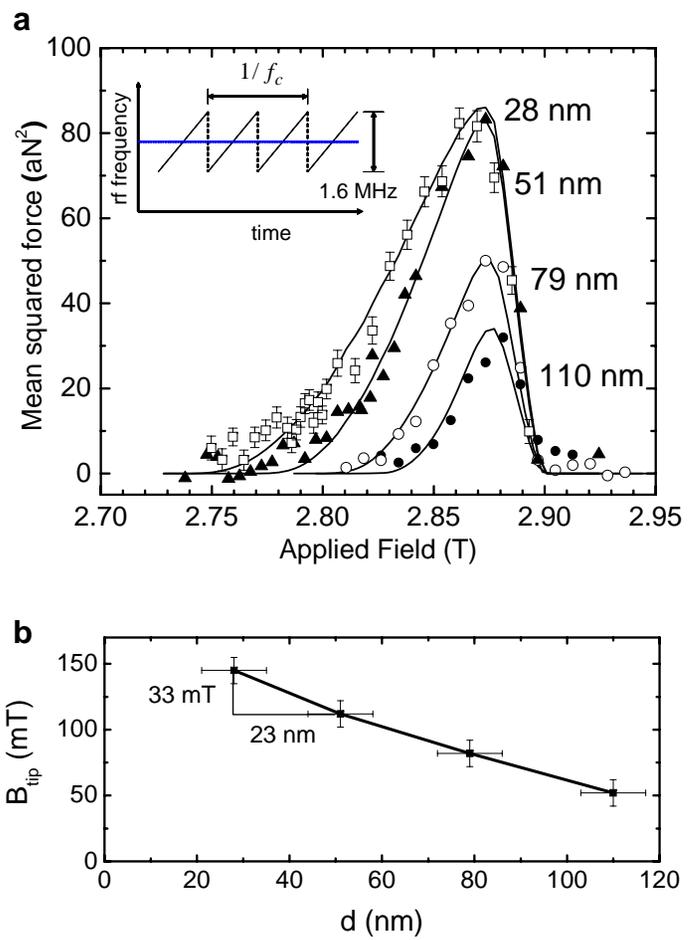

Fig. 2

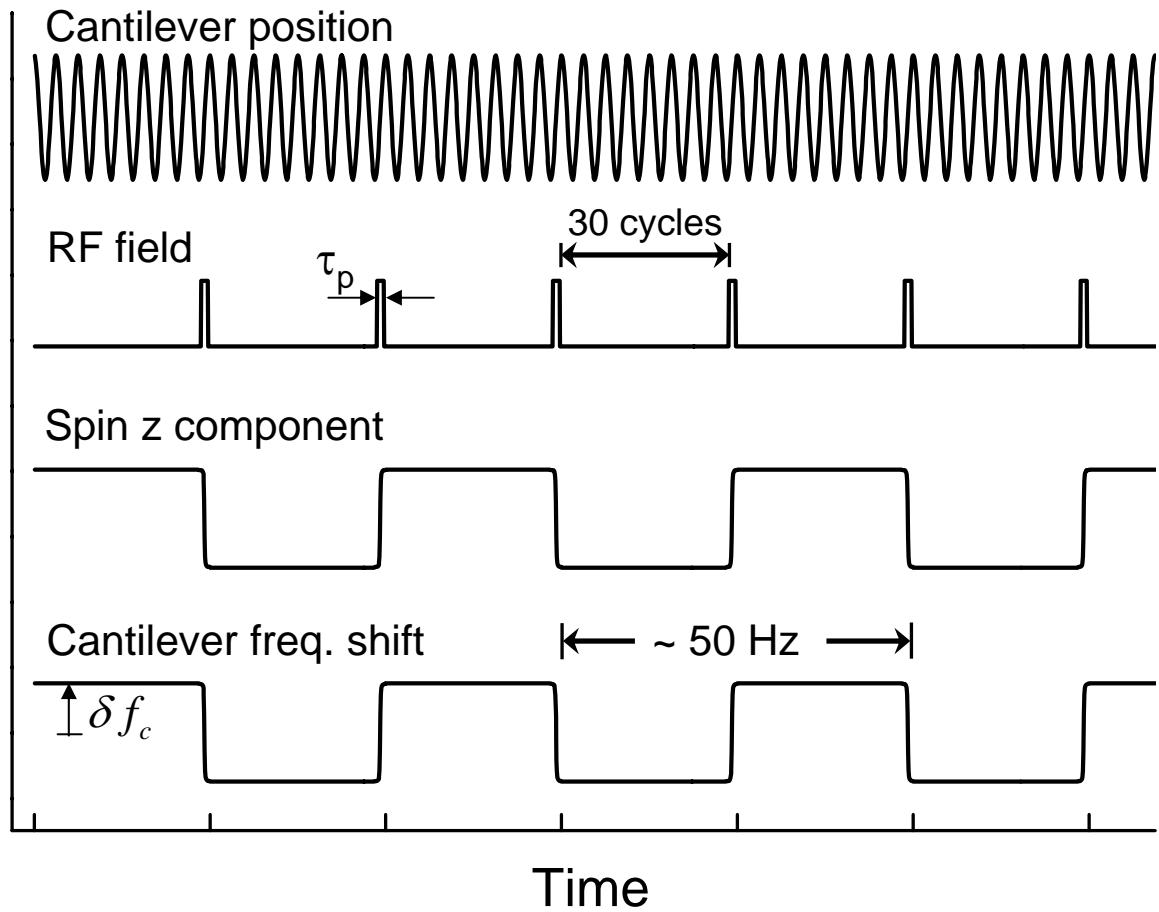

Fig. 3

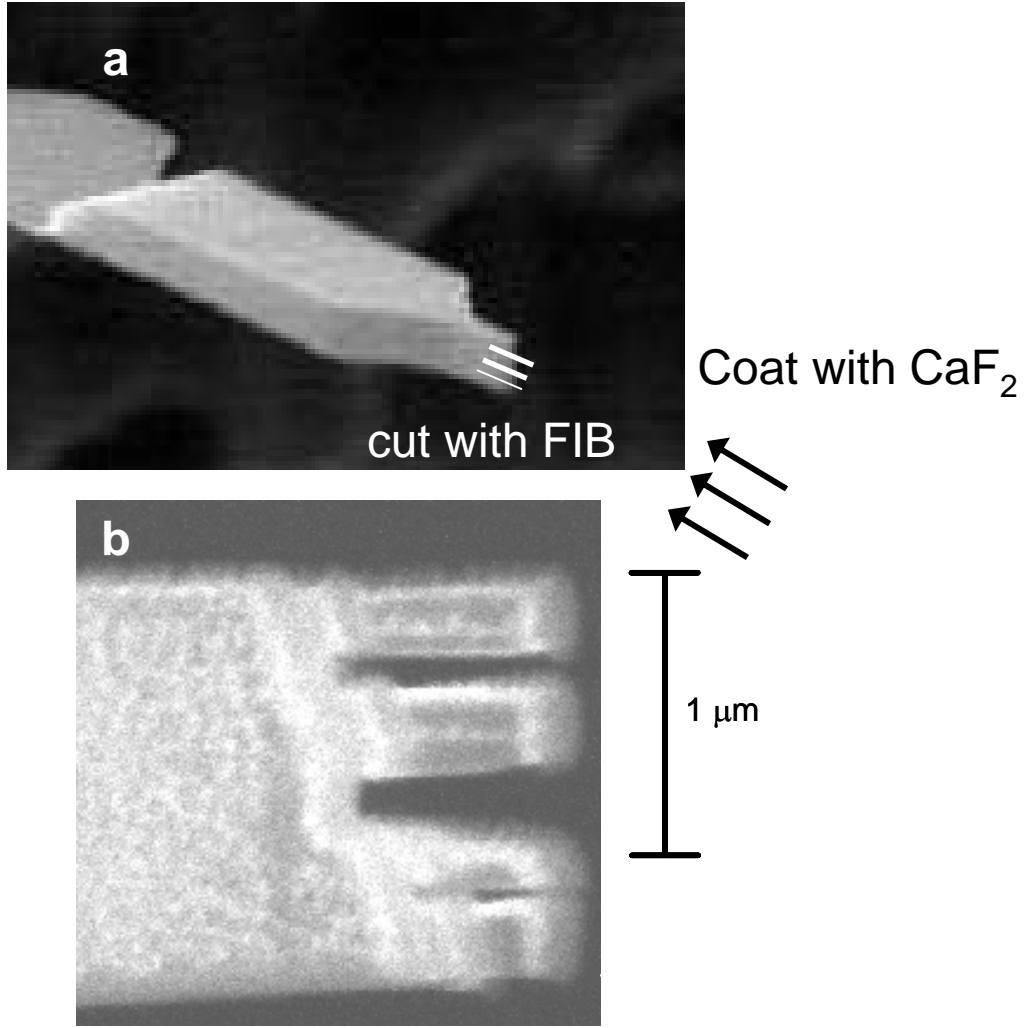

Fig. 4

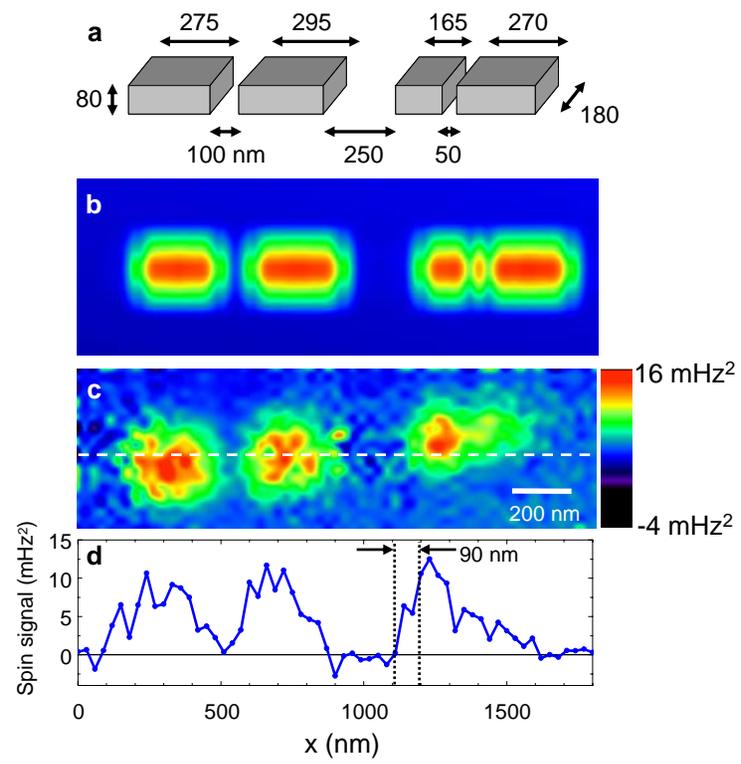

Fig. 5